# Depreciation and the Time Value of Money

Brendon Shaun Farrell - *brendon.farrell@griffithuni.edu.au*

## Abstract

Generally accepted depreciation methods do not compute the intrinsic value of an asset, as they do not factor for the Time Value of Money, a key principle within financial theory. This is disadvantageous, as knowing the intrinsic value of an asset can assist with making effective purchase and sale decisions.

By applying the Time Value of Money principle to deprecation and book valuation, methods can be formulated to approximate the intrinsic valuation of a depreciable asset, which improves the capacity for buyers and sellers of assets to make rational decisions.

A deprecation method is formulated within, which aims to better match book value with intrinsic value. While this method makes many assumptions and thus has limitations, more complex formulas, which factor for a greater number of variables, can be created using a similar approach, to produce better approximations for intrinsic value.

## Introduction

The Time Value of Money (TVM) is a principle within financial theory which assumes individuals prefer to receive money sooner rather than later. TVM is the basis of many financial asset valuation methodologies, which utilise a discount rate to convert expected future cash flows into a single present value equivalent (DeFusco 2004).

These methodologies do not determine the market value of a financial asset, but rather the intrinsic value, which is the value of owning the asset for the entirety of its lifetime. When compared with the market value of an asset, the intrinsic value is a useful indicator of whether a potential buyer or seller should proceed with their actions.

While there are numerous well accepted intrinsic valuation methodologies for financial assets, there are none for depreciable assets. Currently, all the deprecation methods accepted by the International Accounting Standards Board disregard the TVM principle (FASB 2016). Thus, the book value of a depreciable asset is not a reliable indicator of intrinsic value and should not be used in making purchasing or sales decisions.

# Approach Overview

It may be thought that the Time Value of Money principle cannot be applied to the valuation of an asset which does not directly yield cash flows in the same manner as a financial asset. However, the intrinsic value of an asset can also be found in any cost savings that result from not having to immediately acquire a replacement.

Consider a business operating as a going concern. Any depreciable asset that is required for operations, needs to be replaced at the end of the asset's life. Once the replacement of the original asset reaches the end of its life, it too shall be replaced, as will those assets which succeed it. This cycle of replacement results in periodic cash outflows, which can be discounted to single cost in present terms using the TVM principle.

As an asset ages, these periodic cash outflows draw closer, reducing the time over which they can be discounted, which in turn results in an inflating present cost. This increase in present cost, which occurs because of an asset's aging, is effectively an asset's intrinsic depreciation expense. Using this logic, it is possible to formulate alternative methods of deprecation, which are superior in terms of their ability to assist decision making.

# Key Assumptions

It is important to recognise, that many factors will affect the intrinsic value of an asset as it ages over time. Several assumptions have been made, in order to demonstrate in a simplistic manner, that it is possible to formulate a deprecation methodology, which utilises the Time Value of Money principle. These assumptions are below:

- Each asset replacement will incur the same cost.
- The owners' cost of capital is constant.
- The asset's productivity does not degrade over its lifetime.
- The total lifetime of the asset and its replacements will be the same.
- The asset will have no residual value at the end of its lifetime.
- The asset does not require any maintenance or repairs during its lifetime.
- A replacement asset will be equally productive.
- The venture requiring the asset will operate indefinitely.
- The asset will be immediately replaced upon the end of its lifetime.
- Payment for the replacement will be made upon replacement.
- It is appropriate for compounding to occur periodically rather than continuously.
- No taxation.

# PRESENT COST OF OWNERSHIP

Determining a formula to compute the intrinsic value of an asset, can be approached by firstly determining the present cost of all cash outflows associated with the purchase of an asset and its subsequent replacements. When the purchase of an asset is imminent, the present expense of indefinite asset ownership is equal to the below formula:

$$P = C + \sum_{k=1}^{\infty} \frac{C}{(1+r)^{lk}}$$

*Where:*
**P** = *Present cost of asset ownership*
**C** = *Cost of replacement (a negative value)*
**r** = *Cost of capital*
**l** = *Lifetime of the asset*
**k** = *Replacement count*

To simplify, both sides of the formula are multiplied by the discount factor:

$$P(1+r)^l = C(1+r)^l + C + \sum_{k=1}^{\infty} \frac{C}{(1+r)^{lk}}$$

Then the difference between the multiple and the original is determined, to cancel the large operators:

$$P(1+r)^l - P = C(1+r)^l + C + \sum_{k=1}^{\infty} \frac{C}{(1+r)^{lk}} - C - \sum_{k=1}^{\infty} \frac{C}{(1+r)^{lk}}$$

Allowing a simplified version of the original formula to be determined:

$$P(1+r)^l - P = C(1+r)^l$$

$$P((1+r)^l - 1) = C(1+r)^l$$

$$P = \frac{C(1+r)^l}{(1+r)^l - 1}$$

# INTRINSIC VALUATION MODEL

If an asset has some portion of remaining lifetime, then it delays all the repurchases by this timeframe. The delayed present expense of indefinite asset ownership, can be computed by the below formula, which discounts the entire perpetuity according to the delay:

$$D = \frac{P}{(1+r)^{l-a}}$$

*Where:*
***D**= Delayed present expense of indefinite asset ownership*
***a** = Age of the asset*

Which can be expanded and simplified:

$$D = \frac{C(1+r)^l}{(1+r)^l - 1} \times \frac{1}{(1+r)^{l-a}}$$

$$D = \frac{C(1+r)^l}{(1+r)^l - 1} \times \frac{(1+r)^a}{(1+r)^l}$$

$$D = \frac{C(1+r)^a}{(1+r)^l - 1}$$

As the value of an asset is equivalent to the effective saving that result from ownership, it can be computed by subtracting the Present Expense from the Delayed Present Expense:

$$V = D - P$$

*Where:*
***V**= Intrinsic Value of the asset*

The expense formulas can now be substituted into the above formula and simplified to determine the Intrinsic Valuation Model:

$$V = \frac{C(1+r)^a}{(1+r)^l - 1} - \frac{C(1+r)^l}{(1+r)^l - 1}$$

$$V = C \times \frac{(1+r)^a - (1+r)^l}{(1+r)^l - 1} \blacksquare$$

# INTRINSIC DEPRECIATION

The intrinsic depreciation expense incurred over a period, is equivalent to the intrinsic value of the asset at the end of the period, less the asset's intrinsic value at the start of the period:

$$\Delta V = V_{end} - V_{start}$$

*Where:*
*$\Delta V$ = The intrinsic depreciation expense incurred over the period*
*$V_{end}$ = The intrinsic value of the asset at the end of the period*
*$V_{start}$ = The intrinsic value of the asset at the start of the period*

Which can be expanded into:

$$\Delta V = C \times \frac{(1+r)^{a_{end}} - (1+r)^l}{(1+r)^l - 1} - C \times \frac{(1+r)^{a_{start}} - (1+r)^l}{(1+r)^l - 1}$$

Where:
*$a_{end}$ = The age of the asset at the end of the period*
*$a_{start}$ = The age of the asset at the start of the period*

Then simplified:

$$\Delta V = C \times \left( \frac{(1+r)^{a_{end}} - (1+r)^l}{(1+r)^l - 1} - \frac{(1+r)^{a_{start}} - (1+r)^l}{(1+r)^l - 1} \right)$$

$$\Delta V = C \times \frac{\left((1+r)^{a_{end}} - (1+r)^l\right) - \left((1+r)^{a_{start}} - (1+r)^l\right)}{(1+r)^l - 1}$$

$$\Delta V = C \times \frac{(1+r)^{a_{end}} - (1+r)^l - (1+r)^{a_{start}} + (1+r)^l}{(1+r)^l - 1}$$

$$\Delta V = C \times \frac{(1+r)^{a_{end}} - (1+r)^{a_{start}}}{(1+r)^l - 1} \blacksquare$$

# COMPARISON OF DEPRECIATION METHODS

Different deprecation methods, will result in different values for depreciable assets. Below is a comparison of depreciation methods applied to an asset with a replacement value of $100 and a lifetime of 10 years; specifically, Intrinsic ($r$ = 20%), Straight-line, Double Declining Balance and Sum of Years Digits.

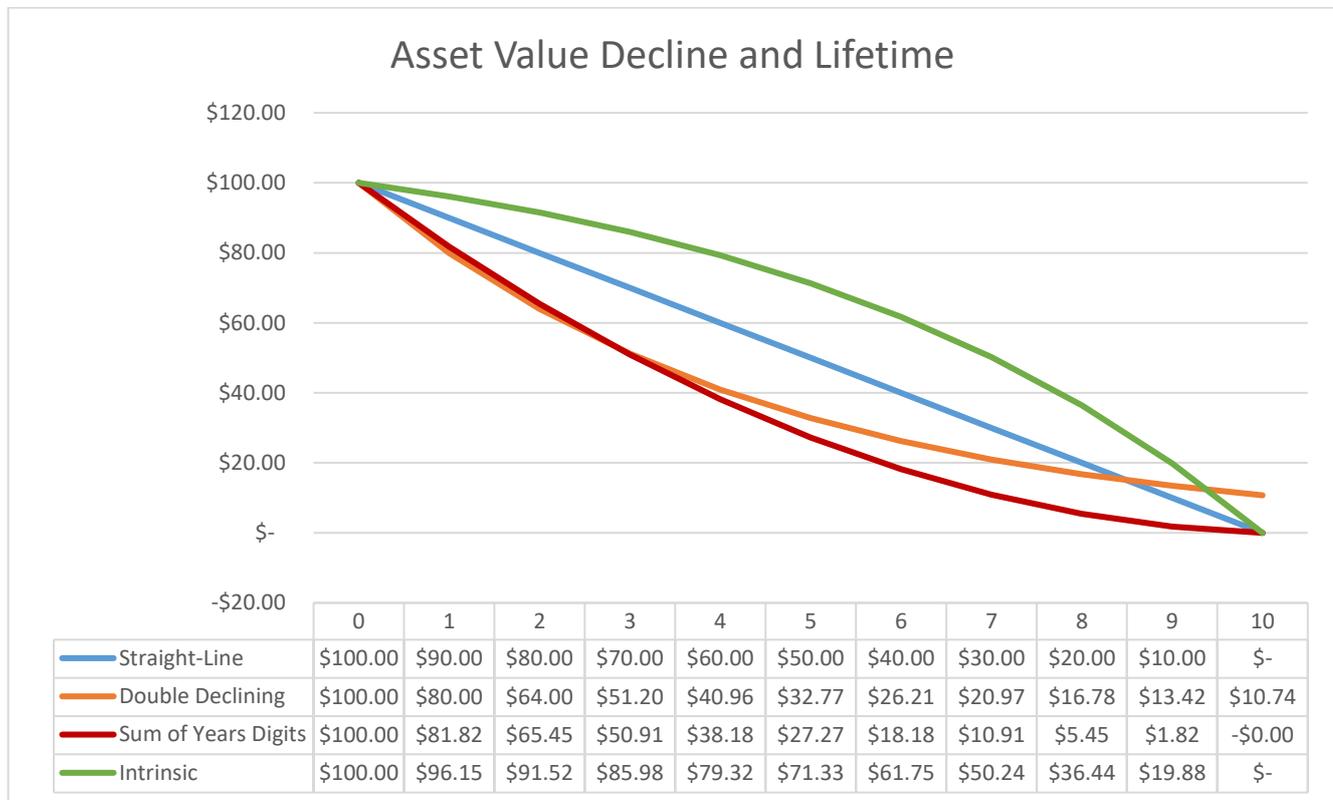

| | 0 | 1 | 2 | 3 | 4 | 5 | 6 | 7 | 8 | 9 | 10 |
|---|---|---|---|---|---|---|---|---|---|---|---|
| Straight-Line | $100.00 | $90.00 | $80.00 | $70.00 | $60.00 | $50.00 | $40.00 | $30.00 | $20.00 | $10.00 | $- |
| Double Declining | $100.00 | $80.00 | $64.00 | $51.20 | $40.96 | $32.77 | $26.21 | $20.97 | $16.78 | $13.42 | $10.74 |
| Sum of Years Digits | $100.00 | $81.82 | $65.45 | $50.91 | $38.18 | $27.27 | $18.18 | $10.91 | $5.45 | $1.82 | -$0.00 |
| Intrinsic | $100.00 | $96.15 | $91.52 | $85.98 | $79.32 | $71.33 | $61.75 | $50.24 | $36.44 | $19.88 | $- |

It can be observed that accelerated and linear depreciation methods such as Straight-line, Double Declining Balance and Sum of Years Digits, start off with higher depreciation rates than Intrinsic. Subsequently, these popular methods either maintain the same depreciation rate or it decreases over time, whist Intrinsic's rate increases.

Due to the differing deprecation rates, Intrinsic computes a higher book value during asset life, than accelerated and linear methods [*Except Double Declining Balance, which fails to fully depreciate the asset in its final year*] but will still write-off the asset's entire cost by the end of its final year.

# COST OF CAPITAL VARIATION

Because the Intrinsic method is underpinned by the Time Value of Money principle, the venture's cost of capital is a key factor in determining intrinsic value. The impact on intrinsic value caused by cost of capital variation, can be observed in the graph below; where an asset with a replacement value of $100, is depreciated over its 10-year lifetime, using the Intrinsic Method at varying costs of capital.

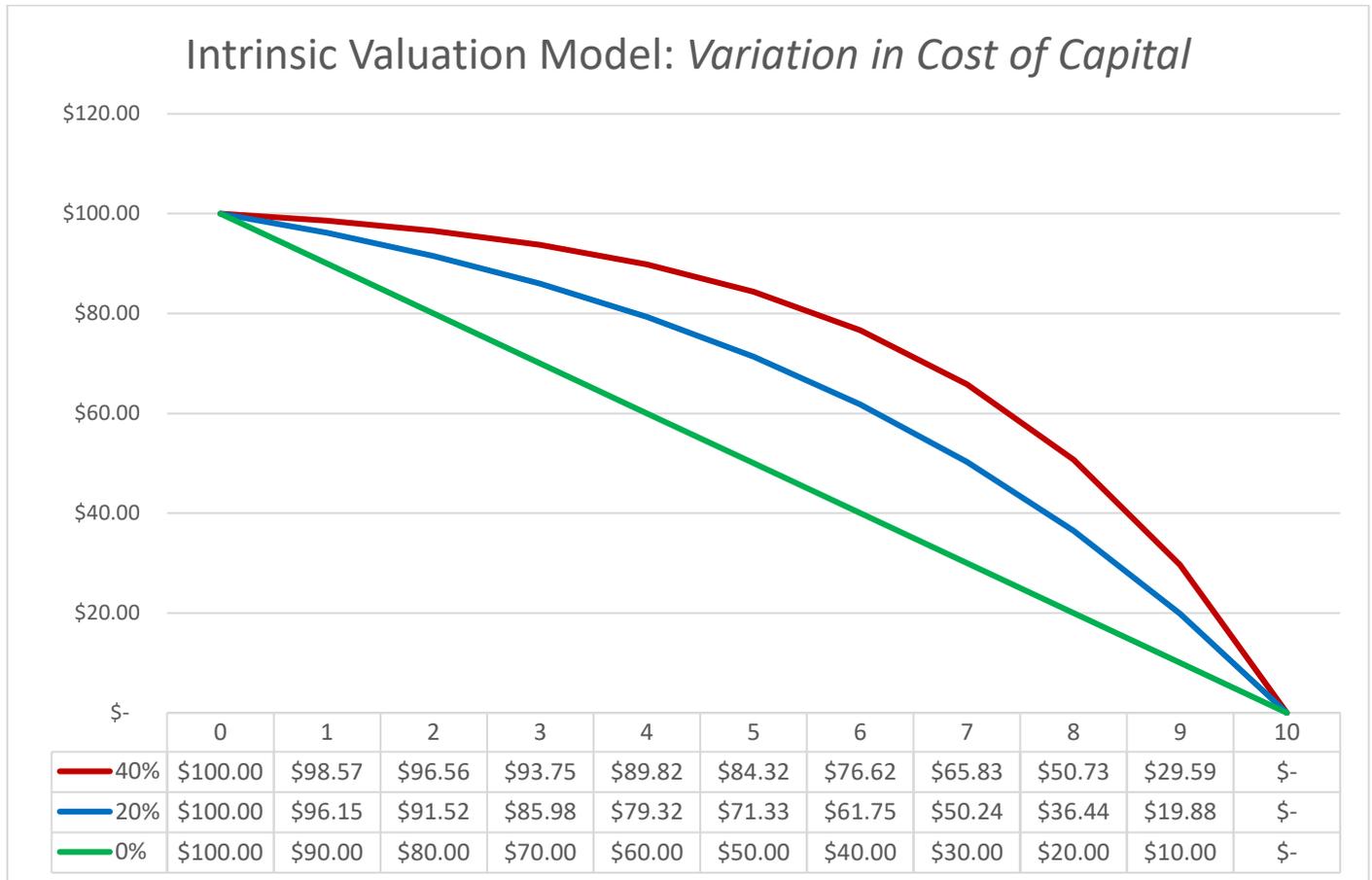

**Intrinsic Valuation Model:** *Variation in Cost of Capital*

| | 0 | 1 | 2 | 3 | 4 | 5 | 6 | 7 | 8 | 9 | 10 |
|---|---|---|---|---|---|---|---|---|---|---|---|
| 40% | $100.00 | $98.57 | $96.56 | $93.75 | $89.82 | $84.32 | $76.62 | $65.83 | $50.73 | $29.59 | $- |
| 20% | $100.00 | $96.15 | $91.52 | $85.98 | $79.32 | $71.33 | $61.75 | $50.24 | $36.44 | $19.88 | $- |
| 0% | $100.00 | $90.00 | $80.00 | $70.00 | $60.00 | $50.00 | $40.00 | $30.00 | $20.00 | $10.00 | $- |

It can be observed that higher costs of capital initially result in lower rates of depreciation, but change to higher rates when the asset nears the end of its lifetime. Thus, a higher cost of capital results in a higher average asset value throughout its lifetime. When graphed, the effect of differing cost of capital is most notable in the convexity of the value curve. Higher discount rates result in greater convexity, whereas discount rates approaching 0% will result in near straight lines.

# Caveats

Because assumptions were made to ensure the Intrinsic Valuation Model within had minimal variables, it will only be accurate in limited circumstances. An example of such a circumstance might be an asset that functions normally throughout its life, but is replaced because it is legally mandated. [*i.e. a critical part inside an aircraft that is replaced for safety reasons*].

This basic form of an Intrinsic Valuation Model has less utility when it comes to assets such as commercial vehicles, which often have a scrap value, require both routine and non-routine repairs, may decrease in utility over time and might have several disadvantages over more technologically advanced replacements.

In instances that are not matching with the assumptions made, other depreciation methodologies may incidentally be a superior gauge of intrinsic value. It is expected however, that improved Intrinsic Valuation Models, which factor for more variables, will be more accurate than currently well accepted depreciation methodologies in such instances.

# Implications

The basic Intrinsic Valuation Model, demonstrates that it is possible to utilise the Time Value of Money, to determine the intrinsic value of a depreciable asset. Because TVM is a critical factor in the valuation of many other asset classes, it can be concluded that existing depreciation methodologies, are excessively misrepresenting the intrinsic value of depreciable assets and that Intrinsic deprecation or similar methods should be adopted.

Additionally, it has been demonstrated that a depreciable asset's intrinsic value will vary depending on the cost of capital applied to it. Because of this, market participants with a relatively low cost of capital, may be incentivised to sell assets before the end of their useful life. Buyers with a higher cost of capital, may be willing to pay a relatively high price, for the intrinsic value they place on the asset is inflated. This insight may assist business decision making or explain some behaviours that are observed within used asset markets.